\documentclass{PoS}

\usepackage{amsfonts} 
\usepackage{amssymb} 
\usepackage{amsmath} 
\usepackage{subfigure}
\usepackage{graphicx} 
\usepackage{array} 
\usepackage{dcolumn} 
\usepackage{bm} 
\usepackage{latexsym} 
\usepackage{longtable} 
\usepackage{color}
\usepackage{mathrsfs}
\usepackage{bbold}
\usepackage{multirow}
\usepackage{arydshln} 

\graphicspath{{./figs/}}
\newcommand{\MeV}{\mathop{\rm MeV}\nolimits}
\newcommand{\GeV}{\mathop{\rm GeV}\nolimits}
\newcommand{\fm}{\mathop{\rm fm}\nolimits}

\newcommand{\vect}[1]{\ensuremath{\boldsymbol{#1}}}
\newcommand{\vprp}[1]{\vect{#1}_{\mathrm T}}

\newcommand{\bvec}{b}

\newcommand{\mN}{m_N}

\newcommand{\kei}{k}

\newcommand{\zetahat}{{\hat \zeta}}

\title{ 
  Lattice QCD calculations of nucleon transverse momentum-dependent parton distributions using clover and domain wall fermions
}

\ShortTitle{
  Lattice QCD calculation of nucleon TMDs at 300 MeV pion mass
}

\author{\speaker{B.~Yoon}$^{~a,\dagger}$, T.~Bhattacharya$^a$, M.~Engelhardt$^b$, J.~Green$^c$, R.~Gupta$^a$, P.~H\"agler$^d$, B.~Musch$^d$, J.~Negele$^e$, A.~Pochinsky$^e$, S.~Syritsyn$^f$ \\
  $^a$Los Alamos National Laboratory, MS B283, P.O. Box 1663, Los Alamos, NM 87545, USA \\
  $^b$Department of Physics, New Mexico State University, Las Cruces, NM 88003-8001, USA \\
  $^c$Institut f\"ur Kernphysik, Johannes Gutenberg-Universit\"at Mainz, 55099 Mainz, Germany \\
  $^d$Institut f\"ur Theoretische Physik, Universit\"at Regensburg, 93040 Regensburg, Germany \\
  $^e$Center for Theoretical Physics, Massachusetts Institute of Technology, Cambridge, MA 02139, USA \\
  $^f$Theory Center, Thomas Jefferson National Accelerator Facility, Newport News, VA 23606, USA
  $^\dagger$E-mail: \email{boram@lanl.gov}}

\abstract{
We present a lattice QCD calculation of transverse momentum dependent parton
distribution functions (TMDs) of protons using staple-shaped Wilson lines.
For time-reversal odd observables, we calculate the generalized Sivers and 
Boer-Mulders transverse momentum shifts in SIDIS and DY cases, and for T-even
observables we calculate the transversity related to the tensor charge and the 
generalized worm-gear shift.
The calculation is done on two different $n_f=2+1$ ensembles: domain-wall fermion 
(DWF) with lattice spacing $0.084\fm$ and pion mass of $297\MeV$, and clover fermion
with lattice spacing $0.114\fm$ and pion mass of $317\MeV$.
The results from those two different discretizations are consistent with each other.
}

\FullConference{The 33rd International Symposium on Lattice Field Theory,\\
    14 - 18 July 2015\\
    Kobe International Conference Center, Kobe, Japan}

\begin{document}

\section{Introduction and Methodology} 
Intrinsic motion of partons inside nucleons gives an important picture of the 
nucleon structure.
The intrinsic motion of partons in the transverse plane in momentum space
is described by the transverse momentum-dependent parton distributions (TMDs).
TMDs have been studied actively both theoretically, \emph{i.e.}, by using the
QCD factorization \cite{Ji:2004xq, Ji:2004wu, Aybat:2011zv} , and 
experimentally, \emph{i.e.}, COMPASS, HERMES and JLab experiments 
\cite{Airapetian:2009ae, Adolph:2012sp, Hulse:2015caa}.
Here we report a lattice QCD calculation of the TMDs.

The methodology we use for the calculation of the TMDs is elaborately described in 
the previous papers~\cite{Musch:2010ka, Musch:2011er}.
It can be summarized as follows.
Considering off-light-cone kinematics from the beginning, TMD correlators are 
parametrized in terms of the Lorentz-invariant amplitudes, so that they can be
calculated on the lattice.
Only the ratios of isovector ($u-d$) TMD observables, that cancel soft factors, 
multiplicative renormalization factors and disconnected quark loop contributions, 
are calculated.
The Semi-Inclusive Deep-Inelastic Scattering (SIDIS) and Drell-Yan (DY) scattering 
processes are described by the staple-shape gauge links connecting two separated 
quarks in the TMD correlators, whose direction and extent are denoted by $v$ and 
$\eta$, respectively.
Here the staple direction $v$ is taken to be space-like in order to avoid the
rapidity divergences \cite{Collins:1981uw};
light-cone can be approached in the limit of the Collins-Soper parameter 
$\zetahat \rightarrow \infty$, guided by a perturbative evolution
\cite{Collins:1981uk, Aybat:2011ge}.

In this study, we analyze two ensembles; we call them the Clover and DWF
ensembles.
Lattice parameters are given in Table~\ref{tab:ens}.
Unlike the previous study Ref.~\cite{Musch:2011er}, we use unitary combination 
of sea and valence quarks, \emph{i.e.}, we use the same fermion discretizations 
for the sea and valence quarks.
The two ensembles have different types of fermion discretizations:
one is the RBC/UKQCD lattices with domain-wall fermions at $a=0.084\fm$, 
and the other is the JLab/William\&Mary lattices with clover 
fermions at $a=0.114\fm$.
Both are at approximately the same pion mass, close to $300\MeV$.
By comparing results on those two ensembles, we can see the dependence 
of the TMD observables on discretization effects, including chiral symmetry 
breaking, and scaling violations. 
When we define the TMD ratio observables, we assume that soft factors are 
purely multiplicative, such that they cancel in ratios.
We furthermore assume that the ratios also cancel the quark field 
renormalization factors, so that they are scale invariant.
By comparing the two calculations at different scales, we can see to what 
extent this assumption holds.

\begin{table}[h]
\begin{center}
\begin{tabular}{c|cccccc}
\hline \hline
ID     & Fermion Type & Geometry         & $a (\fm)$ & $m_\pi (\MeV)$ & \# confs. & \# meas. \\
\hline
Clover & Clover       & $32^3 \times 96$ & 0.114     & 317            & 967       & 23208    \\
DWF    & Domain-wall  & $32^3 \times 64$ & 0.084     & 297            & 533       & 4264     \\
\hline \hline
\end{tabular}
\end{center}
\caption{Lattice parameters of the RBC/UKQCD domain-wall ensemble and the 
  JLab/William\&Mary clover ensemble used in this work.
  $n_f = 2+1$ for both ensembles.
}
\label{tab:ens}
\end{table}


\section{Results} 
\label{sec:res}

In this study, we calculate four TMD observables: (1) generalized Sivers
shift, (2) generalized Boer-Mulders shift, (3) transversity $h_1$ and 
(4) generalized worm-gear shift from $g_{1T}$.

The generalized Sivers shift addresses the distribution of unpolarized quarks 
in a transversely polarized proton, and it is defined by
$  \langle \vect{k}_y \rangle_{TU}(\vprp{\bvec}^2;\ldots) 
  \equiv \mN \tilde f_{1T}^{\perp[1](1)}(\vprp{\bvec}^2;\ldots)
  / \tilde f_1^{[1](0)}(\vprp{\bvec}^2;\ldots) \,,$
%
%
where $m_N$ is the nucleon mass, $f_{1T}^{\perp}$ is the Sivers function, 
and $f_1$ is the unpolarized function.
Here tilde represents the Fourier transformation of a function in the 
transverse momentum $\vprp{\kei}$-space onto the $\vprp{\bvec}$-space,
superscript $[1]$ denotes the first Mellin moment in the longitudinal
momentum fraction of the quark, $x$, and superscript $(n)$ are defined
as
$  \tilde f^{(n)}(x, \vprp{\bvec}^2\ldots ) 
  \equiv n!\left( -\frac{2}{\mN^2}\partial_{\vprp{\bvec}^2} \right)^n \ 
  \tilde f(x, \vprp{\bvec}^2;\ldots )$.
%
Detailed descriptions and physical interpretations of the notations are given 
in Ref.~\cite{Musch:2011er}.

\begin{figure*}[tb]
  \includegraphics[width=0.48\linewidth]{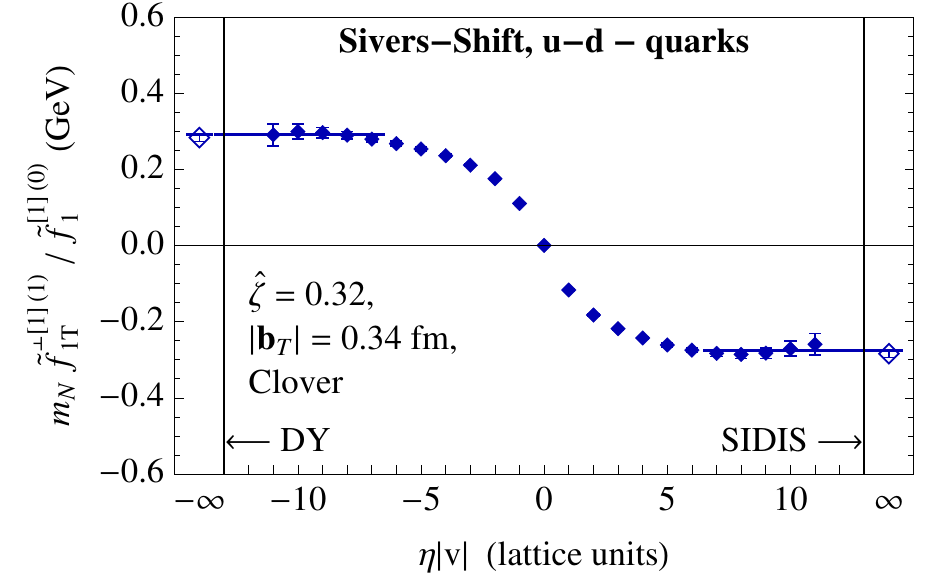}
  \includegraphics[width=0.48\linewidth]{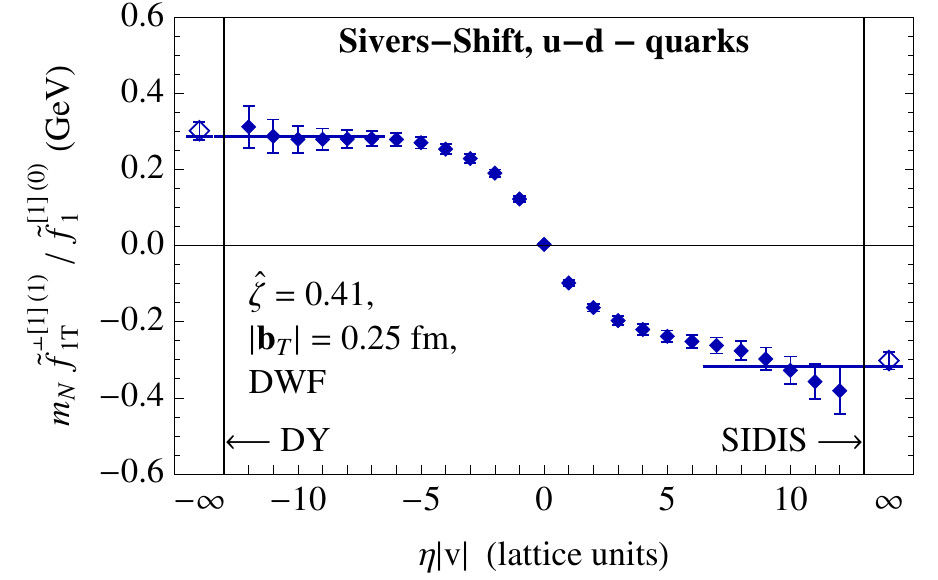}\\\vspace{1em}
  \includegraphics[width=0.48\linewidth]{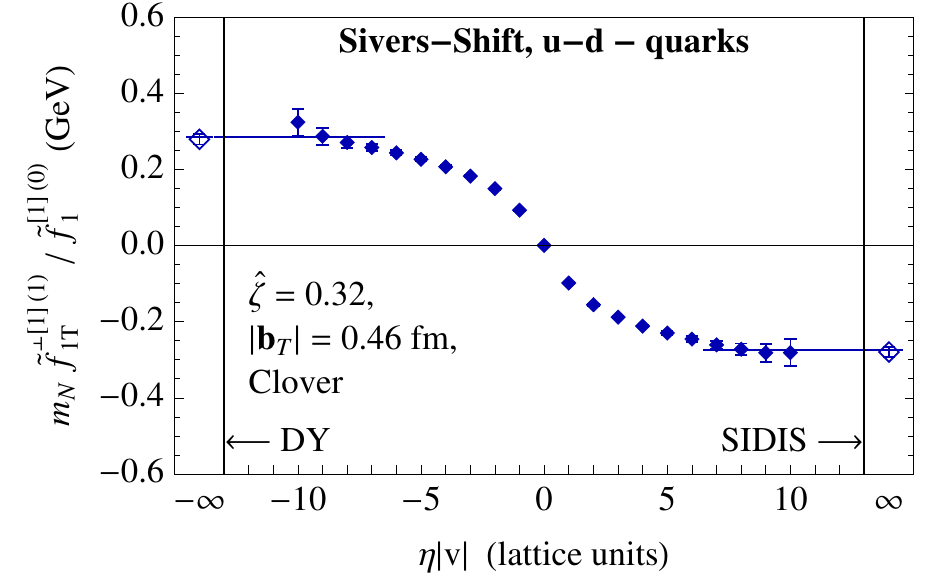}
  \includegraphics[width=0.48\linewidth]{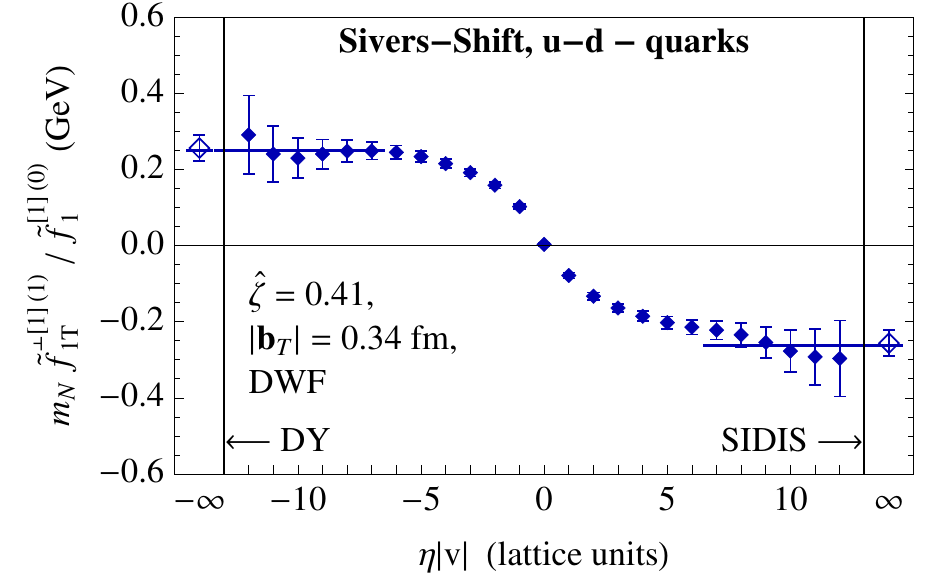}
\caption{
  Dependence of the generalized Sivers shift on the staple extent $\eta|v|$ 
  for Clover (left) and DWF (right) ensembles at $|\vprp{b}|=3a$ (top), and 
  $4a$ (bottom).
  Collins-Soper parameter is fixed at the highest value $\zetahat=0.41$ and 
  $0.32$ for DWF and Clover ensembles, respectively.
}
\label{fig:sivers-etav}
\end{figure*}

In order to obtain the generalized Sivers shift in SIDIS or DY process, the 
staple extent $|\eta||v|$ of the gauge link connecting the two separated 
quarks needs to be extrapolated to infinity.
The $\eta|v|$-dependence of the generalized Sivers shift is shown in 
Fig.~\ref{fig:sivers-etav}.
In our setup, DY process is given by the limit $\eta|v| \rightarrow 
-\infty$, and SIDIS by the limit $\eta|v| \rightarrow \infty$.
As predicted in the QCD factorization, DY and SIDIS results show the opposite
sign and have the same magnitude.
We find that $\eta|v|$-dependence forms a plateau when $|\eta||v| \ge 7a$,
regardless of the lattice spacing.
Hence we obtain the results in DY and SIDIS processes by taking average over
$|\eta||v| = 7a \sim \eta_{\text{max}}$, simultaneously in the DY and 
SIDIS limit, imposing anti-symmetric condition in $\eta|v|$.

Although they form a plateau as they lie on a constant within $1\sim 2\sigma$ 
deviation, we find non-zero slope in some cases, which indicates that
there could be remaining systematic error when $|\eta||v|$ is close to $7a$.
As one can see, in addition, the statistical uncertainty increases as the 
staple extent increases.
Hence, if we use weighted average to obtain the asymptotic value, it will be
governed by the results at small $|\eta||v|$, where the systematic error is
relatively larger than other data points.
In order to avoid this problem, we calculate the asymptotic value by using 
only the mean values of the data points, and calculate the statistical error 
by using the Jackknife method.
Here we use different $\eta_{\text{max}}$ for different measurements.
For Clover, $\eta_{\text{max}}$ is 12 up to $|\vprp{b}|=0.23\fm$, 11 for 
$|\vprp{b}|=0.34\fm$, 10 for $|\vprp{b}|=0.45\fm$, and so on.
For DWF, $\eta_{\text{max}}$ is 12 up to $|\vprp{b}|=0.42\fm$, 10 for 
$|\vprp{b}|=0.50\fm$, 9 for $|\vprp{b}|=0.59\fm$, and so on.
Therefore, the results at large $|\vprp{b}|$ region are obtained only
from a few data points $|\eta||v|$ close to $7a$.
Hence, they have smaller statistical error than those at small $|\vprp{b}|$ 
region, and they may have unresolved systematic error because there are too
small number of data points to confirm if we are in an asymptotic region 
that forms plateau or not.
\begin{figure*}[tb]
  \includegraphics[width=0.48\linewidth]{{{UminusD_SiversRat_zetahat-0.35_bdepend_comb}}}\quad
  \includegraphics[width=0.48\linewidth]{{{UminusD_SiversRat_bT-0.34_evolution_comb}}}
\caption{
  Dependence of the generalized Sivers shift on $|\vprp{b}|$ (left) and 
  $\zetahat$ (right) for the two different ensembles.
  In $|\vprp{b}|$-dependence plot $\zetahat$ is fixed around $0.35$, and 
  in $\zetahat$-dependence plot $|\vprp{b}|$ is fixed at $0.34\fm$.
  When $|\vprp{b}|$ is small, the lattice results are supposed to be 
  contaminated by the lattice cutoff effect.
  In $|\vprp{b}|$-dependence plot, the shaded area is the small $|\vprp{b}|$ 
  region where $|\vprp{b}| \le 3a_{\text{DWF}} \approx 0.25\fm$.
}
\label{fig:sivers-evol}
\end{figure*}

\begin{figure*}[tb]
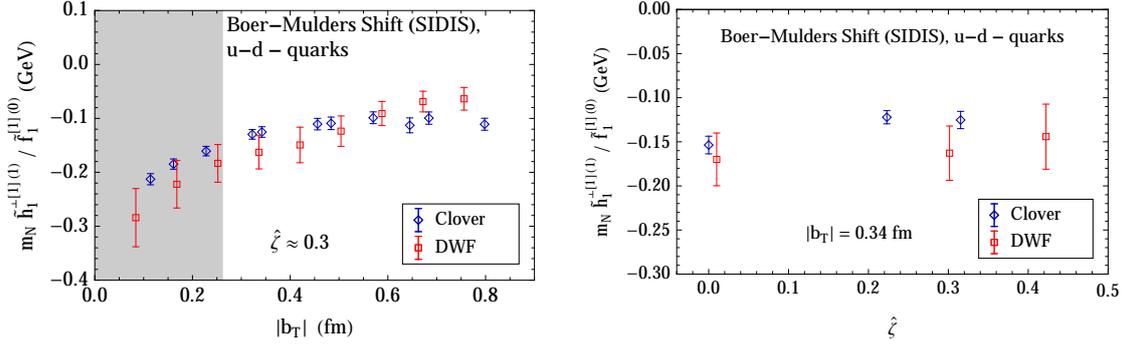

  \includegraphics[width=0.48\linewidth]{{{UminusD_BoerMuldersRat_zetahat-0.35_bdepend_comb}}}\quad
  \includegraphics[width=0.48\linewidth]{{{UminusD_BoerMuldersRat_bT-0.34_evolution_comb}}}
\caption{
  Dependence of the generalized Boer-Mulders shift on $|\vprp{b}|$ (left) and $\zetahat$ 
  (right) for the two different ensembles.
  Notations and other parameters are the same as those in Fig.~\protect\ref{fig:sivers-evol}.
}
\label{fig:boer-evol}
\end{figure*}

\begin{figure*}[tb]
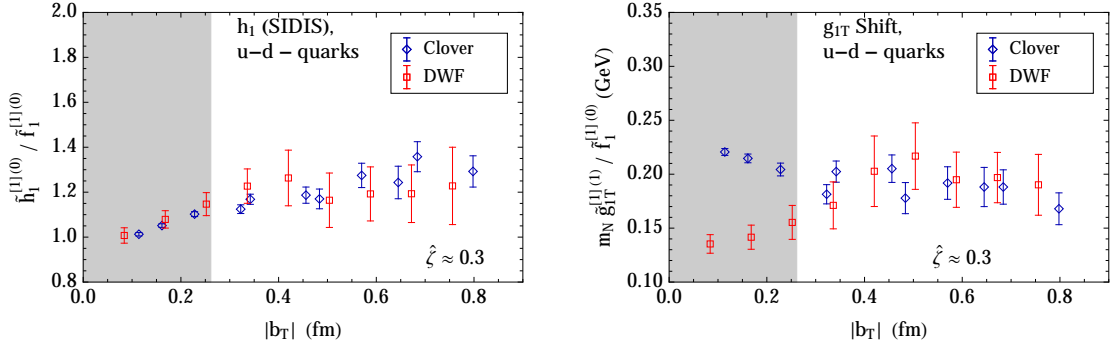

  \includegraphics[width=0.48\linewidth]{{{UminusD_h1Rat_zetahat-0.35_bdepend_comb}}}\quad
  \includegraphics[width=0.48\linewidth]{{{UminusD_g1TRat_zetahat-0.35_bdepend_comb}}}
\caption{
  Dependence of the transversity ratio $\tilde h_{1}^{[1](0)}/\tilde f_1^{[1](0)}$ (left)
  and the generalized $g_{1T}$ (right) on $|\vprp{b}|$ for the two different ensembles.
  Notations and other parameters are the same as those in 
  Fig.~\protect\ref{fig:sivers-evol}.
}
\label{fig:teven-evol}
\end{figure*}

Fig.~\ref{fig:sivers-evol} shows the dependence of the generalized Sivers 
shift on $|\vprp{b}|$ (left) and $\zetahat$ (right) for the two different 
ensembles.
When we define the generalized Sivers shift by the ratio, we assume that the two
quarks in the non-local quark bilinear operators are separated far enough so 
that renormalization constant is independent of the gamma structure of the
operator.
This assumption is correct only when $|\vprp{b}|$ is large enough.
Furthermore, when $|\vprp{b}|$ is small, the lattice results are contaminated
by the lattice cutoff effect.
In $|\vprp{b}|$-dependence plot, the shaded area represents the small 
$|\vprp{b}|$ region where $|\vprp{b}| \le 3a_{\text{DWF}} \approx 0.25\fm$.
Here the minimum separation $3a_{\text{DWF}}$ is chosen so that Clover and 
DWF ensembles of different fermion discretizations and cutoff scales give 
the consistent results for all the $|\vprp{b}|$-dependences of the 
observables given in 
Figs.~\ref{fig:sivers-evol},~\ref{fig:boer-evol}, and \ref{fig:teven-evol}.
%

%
%

The right-hand side plot on Fig.~\ref{fig:sivers-evol} shows the 
Collins-Soper evolution parameter $\zetahat$-dependence of the generalized 
Sivers shift at fixed $|\vprp{b}|=0.34\fm$.
The $\zetahat$ is a scale parameter introduced to avoid the rapidity 
divergences by taking the Wilson line off the light-cone to the
space-like region.
The light-cone limit is $\zetahat\rightarrow\infty$, and TMDs with finite
$\zetahat$ can be evolved to the light-cone perturbatively.
In order to make the perturbative evolution reliable, however, $\zetahat$
needs to be large enough so that $\zeta = 2m_N\zetahat \gg 
\Lambda_{\text{QCD}}$.
In this study, we calculate only up to $\zetahat = 0.41$, which is too small
to use the perturbativative evolution.
The major problem of the lattice QCD calculation at large $\zetahat$ is the
statistical uncertainty; larger $\zetahat$ calculation requires larger 
proton momenta, which introduces large statistical error \cite{Musch:2011er}.
%

%
Another T-odd TMD we calculate is the generalized Boer-Mulders shift, which is 
defined by
  $\langle \vect{k}_y \rangle_{UT}(\vprp{\bvec}^2;\ldots) 
  \equiv \mN {\tilde h_{1}^{\perp[1](1)}(\vprp{\bvec}^2;\ldots)}
  / {\tilde f_1^{[1](0)}(\vprp{\bvec}^2;\ldots)} \,,$
%
which describes the distribution of transversely polarized quarks in an 
unpolarized proton.
Here $h_{1}^{\perp}$ is the Boer-Mulders function.
%

%

%

The dependence of the Boer-Mulders shift on $|\vprp{b}|$ and $\zetahat$ for two
different ensembles is shown in Fig.~\ref{fig:boer-evol}.
We find that the results on the two different Clover and DWF ensembles are 
consistent within their statistical uncertainty.
In the generalized Boer-Mulders shift case, $\zetahat$ dependence has been
studied up to $\zetahat=2.03$ with pions in Ref.~\cite{Engelhardt:2013nba},
exploiting the fact that signal-to-noise ratio is much larger in pions than
in protons.
We expect similar scaling behavior in the proton, but it is not clearly shown
by this study because calculation extends only up to $\zetahat=0.41$.
%


%

For the rest of this section, we briefly describe the results of the T-even TMDs:
the transversity $h_1$ and the generalized worm-gear shift $g_{1T}$.
Unlike the T-odd TMDs, such as the Sivers and Boer-Mulders distributions,
T-even TMDs are predicted to be process independent, \emph{i.e.}, 
same in DY and SIDIS processes.
Hence they have been studied with lattice QCD by using the straight Wilson
line \cite{Hagler:2009mb, Musch:2010ka}, and it has been shown that the
difference between straight gauge link and staple-shaped gauge link is
small for these T-even TMDs \cite{Musch:2011er}.
In this study, we reconfirm this argument for different lattice 
discretizations and lighter pion masses.
We found that most of the results from the two lattice discretizations are 
consistent with each other, except in the small $|\vprp{b}|$ region of the 
generalized worm gear shift $g_{1T}$.
In Fig.~\ref{fig:teven-evol}, we show the $|\vprp{b}|$ dependence of the two
T-even TMDs.
As one can see, the Clover and DWF results of $g_{1T}$ shift are different when 
$|\vprp{b}| \le 0.25\fm$.
However, we expect large systematic uncertainties regarding the lattice 
regularization and renormalization when $|\vprp{b}|$ is small, and the 
results from different discretizations may differ in the region.

\section{Comparison with Phenomenological Estimate and Conclusion}
In this study, we compared the lattice QCD calculation from two different
discretizations: $a=0.114\fm$ Clover vs. $a=0.084\fm$ DWF at $m_\pi \approx 300\MeV$.
We found that the results of the Clover and DWF ensembles are consistent
with each other in the region where lattice artifacts are expected to be
small.
\begin{figure}[tb]
\centering
\includegraphics[width=0.6\textwidth]{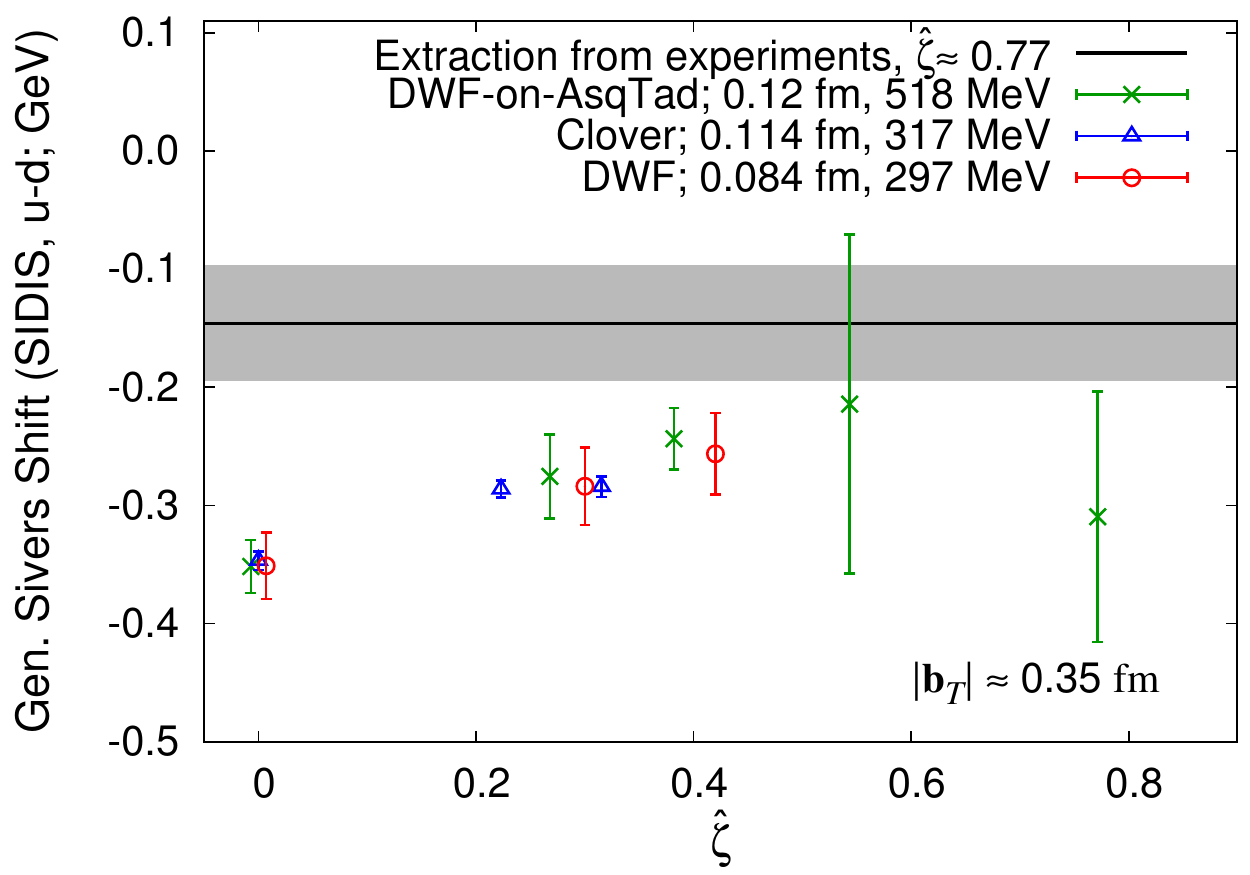}
\caption{ Experimental estimate and lattice QCD calculations of the generalized 
  Sivers shift in SIDIS limit for different Collins-Soper parameter values 
  $\hat{\zeta}$ with various lattice setups.
  The results of DWF-on-Asqtad are obtained from the previous study of 
  Ref.~\cite{Musch:2011er}.
  The extraction from the experiments has been done at $\zetahat\approx 0.77$.
}
\label{fig:comp_exp}
\end{figure}
In Fig.~\ref{fig:comp_exp}, we compare the current lattice QCD calculations
of the generalized Sivers shift with the experimental estimate.
The generalized Sivers shift is defined by the ratio of the Sivers function and 
the unpolarized function.
In order to calculate the experimental estimate of the generalized Sivers shift,
we use the Sivers shift results of HERMES, COMPASS and Jefferson Lab experiments 
extracted in Ref.~\cite{Echevarria:2014xaa}.
In the comparison plot, it is clearly shown that the three lattice ensembles with 
different pion mass and different discretization give consistent results.
One also can find that the discrepancy between the central values of phenomenological 
value and lattice results becomes smaller as $\zetahat$ increases.
Here the $\zetahat$ is a scale introduced to regulate the rapidity divergences,
and the factorization works only when $\zetahat$ is large enough so that it
lies in the perturbative regime. 
The extraction from the experiments has been done at $\zetahat\approx 0.77$,
which corresponds to $\zeta \approx 1.5\GeV$, while precise lattice results are 
obtained at $\zetahat \le 0.41$, and the lattice data points at $\zetahat > 0.41$
have huge uncertainty.
Going by the behavior we see in the pion study \cite{Engelhardt:2013nba}, we
expect that extrapolating the lattice results to $\hat{\zeta} \rightarrow 0.77$ 
brings the lattice result close to the extracted value from the experiments.

\section*{Acknowledgments}
Computations were performed using resources provided by the U.S. DOE
Office of Science through the National Energy Research Scientific
Computing Center (NERSC), a DOE Office of Science User Facility, under
Contract No. DE-AC02-05CH11231, as well as through facilities of the
USQCD Collaboration, employing the Chroma software suite~\cite{Edwards:2004sx}. The
RBC/UKQCD collaboration is gratefully acknowledged for providing gauge
configurations analyzed in this work, as are K. Orginos (supported by
DOE grant DE-FG02-04ER41302) and the Jefferson Lab lattice group
(supported by DOE grant DE-AC05-06OR23177, under which Jefferson
Science Associates, LLC, operates Jefferson Laboratory). Support by
the Heisenberg-Fellowship program of the DFG (P.H.),
the PRISMA Cluster of Excellence at the University of Mainz (J.G.),
and the RIKEN Foreign Postdoctoral Researcher Program (BNL) as well as
the Nathan Isgur Fellowship (JLab) (S.S.) is acknowledged. This work
was furthermore supported by the U.S. DOE and the Office of Nuclear
Physics through grants DE-FG02-96ER40965 (M.E.), DE-SC0011090 (J.N.)
and DE-FC02-06ER41444 (A.P.). R.G., T.B. and B.Y. are supported by
DOE grant DE-KA-1401020 and the LDRD program at LANL.

\end{document}